\documentstyle[10pt,epsf,epsfig,hangcaption,xspace,amssymb,amsfonts,amsmath,amsthm,cite,
dp_delphititle,lineno]{dp_delphi}

\makeindex
\def\DpPaperGroup{EP}
\def\DpPaperRef{2002-069}
\def\DpDate{{28 June 2002}}
\def\DpTitle{{ Four-fermion simulation at LEP2 in DELPHI}}
\def\DpSubmit{}
\def\DpComment{ }
\def\DpEMail{ }

\include{newcom}

\begin{document}
\makeatletter
\input{coll.sty}
\makeatother

\begin{titlepage}
\pagenumbering{roman}

\CERNpreprint{\DpPaperGroup}{\DpPaperRef}   
\date{{\small\DpDate}}                      
\title{\DpTitle}                            

\def\thefootnote{*}

\begin{center}
  \vspace*{2.cm} \normalsize {
    {\bf A.~Ballestrero$^{1,2}$ \footnote
    { Work supported in part by the European Union under 
      contract HPRN-CT-2000-00149}, R.~Chierici$^1$, F.~Cossutti$^3$ and 
       E.~Migliore$^{2,4}$} \\ \vskip 0.5 cm
    {\footnotesize \it $^1$ CERN, CH-1211 Geneva, Switzerland.}\\
    {\footnotesize \it $^2$ INFN, Sezione di Torino, IT-10125 Torino, Italy.}\\
    {\footnotesize \it $^3$ INFN, Sezione di Trieste, IT-34127 Trieste,
    Italy.}\\
    {\footnotesize \it $^4$ Dipartimento di Fisica Sperimentale, Universit\`{a}
      di Torino, IT-10125 Torino, Italy.}\\     }
\end{center}

\begin{shortabs}                            
\noindent
We present and discuss the generator setup for
$e^+e^-\rightarrow 4f$ processes chosen by the DELPHI collaboration.
The need to combine the most recent theoretical achievements in the
CC03 sector with the state of the art description of the remaining
part of the 4-fermion processes has led to an original combination of
different codes, with the {\tt WPHACT 2.0} 4-fermion generator and the
{\tt YFSWW} code for the CC03 $\mathcal{O}(\alpha)$ corrections as a
starting point.  The coverage of the 4-fermion phase space is
discussed in detail, with particular attention to ensuring the
compatibility of {\tt WPHACT} with dedicated $\gamma\gamma$ generators.
\end{shortabs}

\vfill

\begin{center}
\DpSubmit \ \\          
\DpComment \ \\
\DpEMail \ \\
\end{center}

\vfill
\clearpage

\headsep 10.0pt

\addtolength{\textheight}{10mm}
\addtolength{\footskip}{-5mm}
\begingroup
%
\newcommand{\DpName}[2]{\hbox{#1$^{\ref{#2}}$},\hfill}
\newcommand{\DpNameTwo}[3]{\hbox{#1$^{\ref{#2},\ref{#3}}$},\hfill}
\newcommand{\DpNameThree}[4]{\hbox{#1$^{\ref{#2},\ref{#3},\ref{#4}}$},\hfill}
\newskip\Bigfill \Bigfill = 0pt plus 1000fill
\newcommand{\DpNameLast}[2]{\hbox{#1$^{\ref{#2}}$}\hspace{\Bigfill}}

%
\footnotesize
\noindent
\normalsize
\endgroup

\addtolength{\textheight}{-10mm}
\addtolength{\footskip}{5mm}
\clearpage

\headsep 30.0pt
\end{titlepage}

%
\pagenumbering{arabic}                              
\setcounter{footnote}{0}                            %
\large
\section{Introduction}
\label{introduction}


This report presents the description of the generator setup which has
been developed for the simulation of four-fermion (4-$f$) processes at LEP, 
and used by the DELPHI experiment for the final LEP2 analyses.

A large variety of first order Feynman diagrams contribute to the
production of four fermions in $e^+e^-$ interactions, depending on the
specific final state.  The possible processes are divided into 3
classes: charged current (CC), neutral current (NC), and mixed current
(MIX), the last receiving contributions from both charged and
neutral current diagrams.  The detailed classification of the 4-$f$
final states used throughout this paper is the one usually adopted at
LEP2~\cite{lep2}, and can be found for instance in~\cite{4fclass}. 
All the 4-$f$ final states described in this classification are listed in
appendix B.
The different families of Feynman diagrams contributing to the production
of 4-$f$ final states at tree level are presented in~\cite{4fdiag};
they are conventionally classified as either $t$-channel or
$s$-channel diagrams, the former referring to diagrams with at least
one boson propagator in the $t$-channel. Thus $t$-channel diagrams are
always present for processes with at least one electron (or positron)
in the final state.


The origin of the work described in this paper is in the outcome of
the 2000 LEP2 Monte Carlo workshop~\cite{lep2mcws} and subsequent work
by the theory community. They have addressed the problems posed by the
LEP experiments in the 4-fermion sector, which were basically: 1) to
provide a theoretical precision in the description of the WW physics
significantly better than the anticipated experimental one; 2) to
supply good modelling of the remaining 4-fermion processes, in order
to compare with the LEP measurements and to give a solid description
of the background to the new physics signals searched for at LEP.


The most important request for $WW$ physics concerned the use of
$\mathcal{O}$($\alpha$) radiative corrections in the so-called Double Pole
Approximation (DPA)~\cite{dpa1,dpa2}. For the virtual part these corrections
are applied only to CC03 diagrams, -- those which correspond to production and
decay of the two $W$s. At present, two Monte Carlo codes include the DPA
corrections: RacoonWW~\cite{RacoonWW}, which makes use of the results of
ref.~\cite{dpa2}, and YFSWW~\cite{Yfsww}, which implements the electroweak
corrections to $WW$ production of reference~\cite{dpa3} and the Khoze-Chapovsky
ansatz~\cite{KCansatz} for the non-factorizable corrections. Different
strategies are also used for real part corrections. For a detailed analysis of
the two approaches and their numerical consistency we refer to~\cite{lep2mcws}. 

It has been demonstrated that the more complete inclusion of first order
corrections leads to significant effects in the precision measurements at
LEP2~\cite{fabio}. In particular, their effects on differential distributions
are so important that the inclusion of those computations in the physics
generators at LEP2 is essential. In addition, there are several other features
that are very desirable (or necessary) to have in a 4-$f$ generator when
studying non-CC03 processes as a signal, or when considering them as a
background to new physics searches. They range from a fully massive calculation
over all the phase space to efficient integration in singular regions, the
inclusion of higher order corrections, etc, and they have been either
unavailable or only partially available in the past.

From the experimentalist's point of view, the ideal 4-$f$ generator
which includes all the above-mentioned features did not exist to
our knowledge at the end of the LEP2 Monte Carlo workshop; in general,
generators which were best suited for CC03 physics did not include the
desirable 4-$f$ features, and vice-versa. This posed a serious problem
of matching between the treatment of different phase space regions in
a coherent event generation.

It is important to note that the 4-$f$ processes overlap with the
so-called $\gamma\gamma$ ones. The overlap is entire for the leptonic
$eell$ final state,\footnote{Here and elsewhere where the meaning is clear
we suppress symbols distinguishing fermion from antifermion in the definition of
4-$f$ states.}
and for the direct photon component (i.e. when the
photon behaves as a point-like particle) in the hadronic $eeqq$ final
state.  In the phase space region where the $\gamma\gamma$ processes
dominate, dedicated codes are usually preferred to 4-$f$ generators
both because they contain the description of the resolved photon
component for the $eeqq$ final state and because they provide a more
efficient generation.  The matching of the genuine 4-$f$ part to the
$\gamma\gamma$ dominated one therefore has to be done carefully in
order to allow a complete and smooth coverage of the whole
experimentally accessible phase space.

\section{The DELPHI approach}
\label{approach}

The DELPHI approach consists in interfacing two generators to keep as
many features as possible for both CC03 and non-CC03 physics.  There
are two ways of doing this:
\begin{itemize}
\item {\bf Independent generation}: This consists of two independent
  event generations, one for the CC03 part with
  $\mathcal{O}$($\alpha$) DPA corrections and one for the remaining
  amplitudes.
  The 4-$f$ matrix element squared in DPA approximation,
  $|4f|^2_{DPA}$, can be decomposed in the following way:
\begin{equation}
|4f|^2_{DPA} = |4f|^2-|\mbox{CC03}|^2+|\mbox{CC03}_{DPA}|^2 =
|4f-\mbox{CC03}|^2+Int.+|\mbox{CC03}_{DPA}|^2 \, ,
\label{4fprime}
\end{equation}
where $4f$ is the 4-fermion matrix element without DPA corrections,
while $\mbox{CC03}$ and $\mbox{CC03}_{DPA}$ represent the CC03 part without and with
their inclusion.  In our notation $Int.$ is the interference term
between the CC03 part and the rest.  It is relevant to note that this
term is computed by using CC03 as given by the Improved Born
Approximation (IBA). With an independent generation one can either
neglect this term, thus introducing a systematic effect of particular
importance for processes of the CC20 class which involve electrons in
the final states, or it is possible to include the interference in the
pure 4-$f$ part. In the latter case, however, it is possible that the
event weight defined in this way, i.e. $|4f-\mbox{CC03}|^2+Int.$, becomes
negative in certain regions of the phase space.
\item {\bf Reweighting}: This basically consists in generating the
  whole 4-$f$ phase space with only one generator, then reweighting
  events to account for the DPA corrections. This correction is
  intended to affect only the CC03 part.  The weights are defined as
  ratios of matrix elements squared, as we will describe in the
  following.
\end{itemize}

We have chosen to adopt the second approach, since dedicated studies
have shown that the negative weight problem in the first one can
become quite sizeable, affecting, for instance, as many as 20\% of the
events in the $qqe\nu$ final states, and is therefore difficult to
treat when an unweighted event generation is needed.

The {\tt KandY} concurrent generators combination~\cite{KandY} has shown that
the second approach provides an effective, though approximate, solution to the
problem, implemented by reweighting full 4-$f$ events (in the original approach
generated with {\tt KORALW}~\cite{Koralw}) with the matrix element provided by
the {\tt YFSWW}~\cite{Yfsww} calculation, in which the $\mathcal{O}$($\alpha$)
radiative corrections are calculated in the leading pole approximation. The
structure of the other currently available  DPA Monte Carlo calculation, {\tt
RacoonWW}~\cite{RacoonWW}, with an explicit 4$f + \gamma$
massless matrix element, is technically not suitable for such an approach.

We have chosen {\tt WPHACT 2.0}~\cite{wphact} as general 4-$f$
generator on top of which the {\tt YFSWW} reweighting has been
implemented. It includes practically all the required features
discussed in the introduction:

\begin{itemize}
\item fully massive matrix elements and phase space for all the 4-$f$
  final states;
\item dedicated phase space mappings for the low mass region, with the
  inclusion of $q\bar{q}$ resonance production via the package
  described in ~\cite{Boonekamp}, and for the small scattering angle
  region, thus allowing a reliable integration for both these regions
  where the cross-section is divergent in the massless approximation;
\item use of the {\tt QEDPS}~\cite{qedps} library to generate ISR
  photons with finite transverse momentum, implemented with the $t$
  scale for $t$-channel dominated processes;
\item Fermion Loop corrections in the IFL (Imaginary part) scheme for
  single $W$ processes~\cite{ifl};
\item running of $\alpha_{QED}$;
\item multiple versions of the Coulomb correction for CC03 events,
  including the Khoze-Chapovsky (K-C) screened Coulomb
  ansatz~\cite{KCansatz}, needed for $\mathcal{O}(\alpha)$
  corrections;
\item off-diagonal CKM matrix elements, except $V_{ub}$.
\end{itemize}

In addition, {\tt WPHACT} allows the generation of unweighted events
for any user-specified subset of 4-$f$ final states in a single run
and the computation of matrix elements with predefined subsets of
Feynman diagrams.

In the DELPHI customized version several features have been added,
which will be described in the next paragraphs:

\begin{itemize}
\item the reweighting for the DPA corrections via {\tt YFSWW} has been
  implemented by interfacing the two codes; in order to allow this the
  YFS exponentiated treatment of the ISR in leading logarithm
  approximation at $\mathcal{O}$$(\alpha^3)$ used in {\tt YFSWW} and
  {\tt KORALW} has been ported and interfaced to {\tt WPHACT};
\item the existing interface with the {\tt PYTHIA}~\cite{pythia}
  hadronization library has been extended to include interfaces with
  the radiation, hadronization and decay libraries {\tt
    PHOTOS}~\cite{photos}, {\tt TAUOLA}~\cite{tauola}, {\tt
    ARIADNE}~\cite{ariadne}, {\tt HERWIG}~\cite{herwig} and the low
  mass hadronization package~\cite{Boonekamp};
\item the possibility to compute the matrix element with different
  subsets of Feynman diagrams has been used to produce for each event
  a list of precomputed squared matrix elements at generation level
  for different contributions to be used in reweightings.
\end{itemize}

The coverage of the phase space in terms of generation cuts has been
studied and optimized with the aim of extending it as much as possible
according to the needs of the physics analyses, but keeping under
control the numerical accuracy of the phase space integration and of
the matrix element calculations. An important part of this work has
been the clear definition of the matching with dedicated
$\gamma\gamma$ generators, i.e. of the separation between classes of
events to be generated with the 4-$f$ code and those to be generated
with specific $\gamma\gamma$ calculations.

\section{$\mathcal{O}(\alpha)$ DPA radiative corrections via {\tt YFSWW}
  reweighting}
\label{DPA}

The weight to be used to account for DPA in 4-$f$ events can be
evaluated as a ratio of matrix elements. Using the notation of
equation~(\ref{4fprime}) the event weight is written as:
\begin{equation}
w = \frac{|4f|^2_{DPA}}{|4f|^2} = 
\frac{|4f|^2-|\mbox{CC03}|^2+|\mbox{CC03}_{DPA}|^2}{|4f|^2} = 
1-\frac{|\mbox{CC03}|^2}{|4f|^2}\left ( 1-\frac{|\mbox{CC03}_{DPA}|^2}{|\mbox{CC03}|^2} \right ) \, .
\label{weight}
\end{equation}
In this reweighting procedure the interference term is included,
although computed using CC03 as given by the IBA.  The numerator in
relation~(\ref{weight}) can be rewritten in a more concise form as:
\begin{equation}
|4f|^2_{DPA} = |\mbox{CC03}|^2(1+\delta_{4f}+\delta_{DPA}) \, ,
\label{additive}
\end{equation}
where:
\begin{equation}
\delta_{4f} = \frac{|4f|^2}{|\mbox{CC03}|^2}-1 \; , \;\;\;\;
\delta_{DPA} = \frac{|\mbox{CC03}_{DPA}|^2}{|\mbox{CC03}|^2}-1 \, . 
\end{equation}
This represents the so-called additive approach to the DPA
reweighting.  In this formulation the new 4-$f$ matrix element results
from the CC03 one with the addition of two corrections, one accounting
for the presence of extra diagrams due to the 4-$f$ background and the
other for the radiative corrections.  The advantage of using the
additive approach is that it depends only upon two ratios, namely
$|\mbox{CC03}|^2/|4f|^2$ and $|\mbox{CC03}_{DPA}|^2/|\mbox{CC03}|^2$.  The first term can
be calculated event by event with an IBA 4-$f$ generator, while the
second can be determined from the output of {\tt YFSWW}.  The {\tt
  YFSWW} generator, used as a reweighter, returns the value
$|\mbox{CC03}_{DPA}|^2/|\mbox{CC03}_{K-C}|^2$, i.e.  the DPA matrix element with
respect to the CC03 one which already includes the Coulomb screening
via the Khoze-Chapovsky correction, an effective treatment of the
non-factorizable part of the $\mathcal{O}(\alpha)$ correction. The
desired ratio can be determined by multiplying this output by
$|\mbox{CC03}_{K-C}|^2/|\mbox{CC03}|^2$, which gives a very small smearing to the
weight distribution, of the order of a 
per cent.

The reader should bear in mind that the DPA correction can be
considered by definition to be reliable only within a few
$\Gamma_{W}$ around the double resonant pole.

\begin{figure}[hbtp]
\begin{center}
  \(
  \begin{array}{cc}
    \scalebox{0.4}{\includegraphics{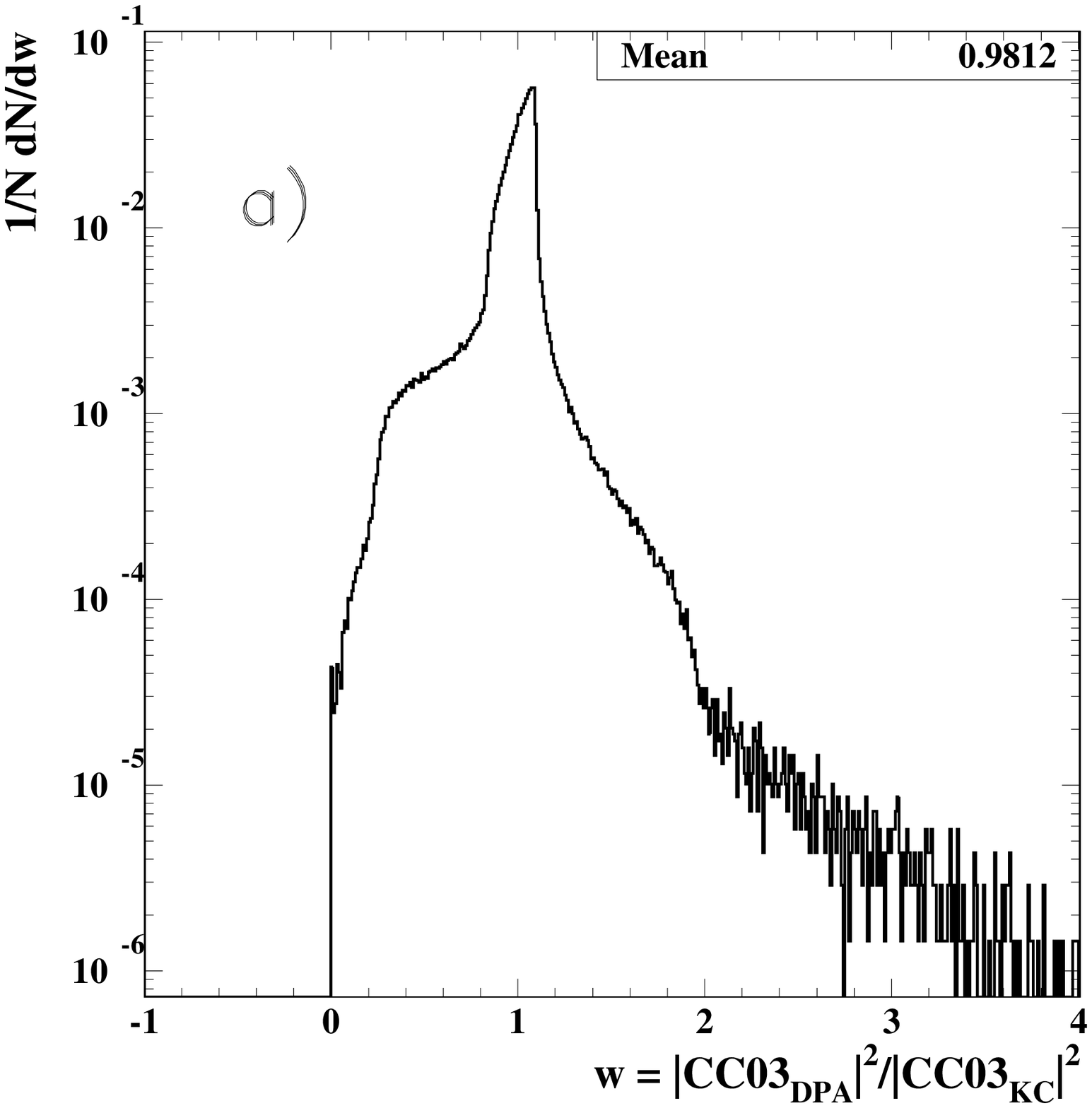}}
    \scalebox{0.4}{\includegraphics{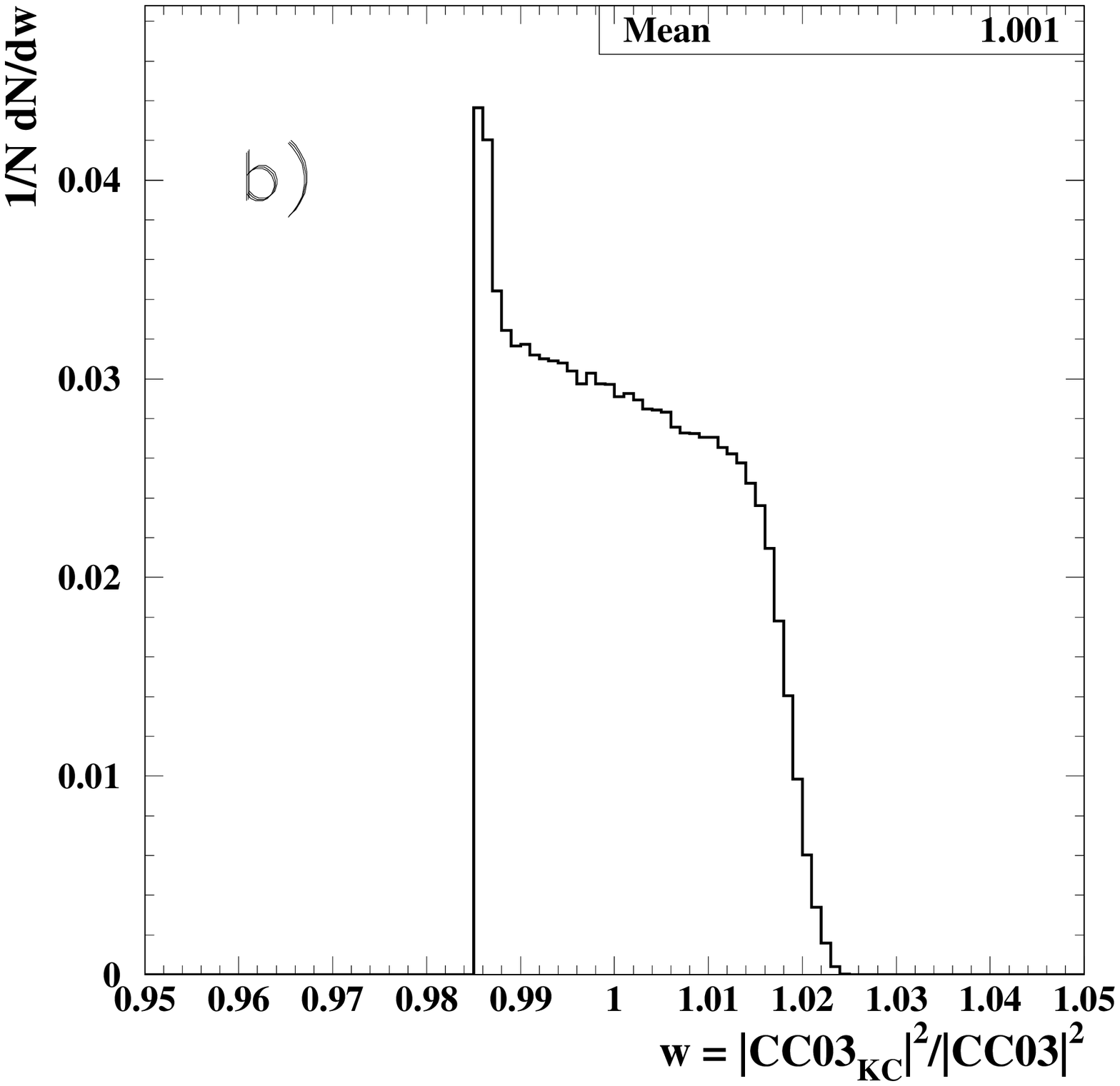}} \\
    \scalebox{0.4}{\includegraphics{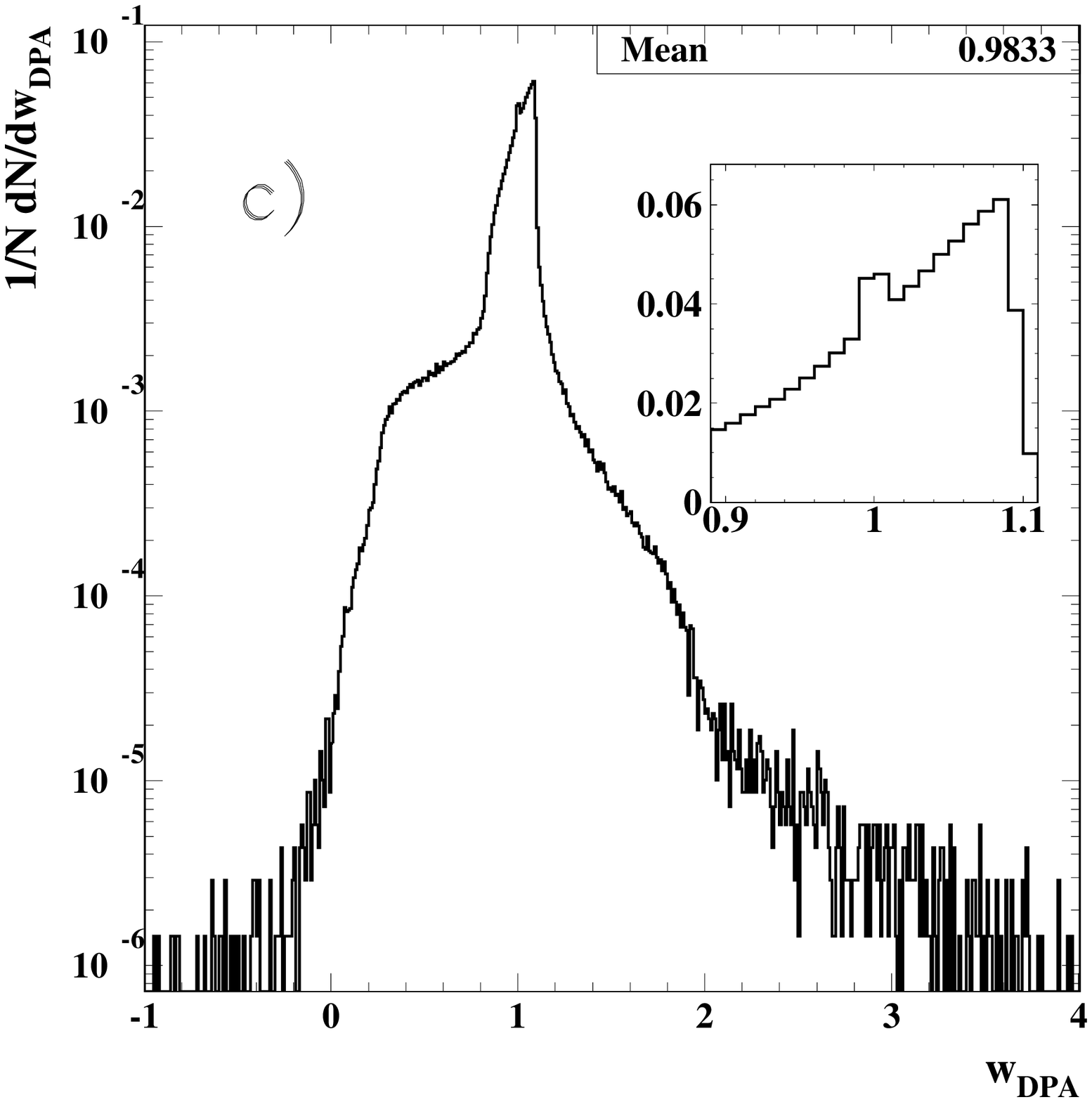}}
    \scalebox{0.4}{\includegraphics{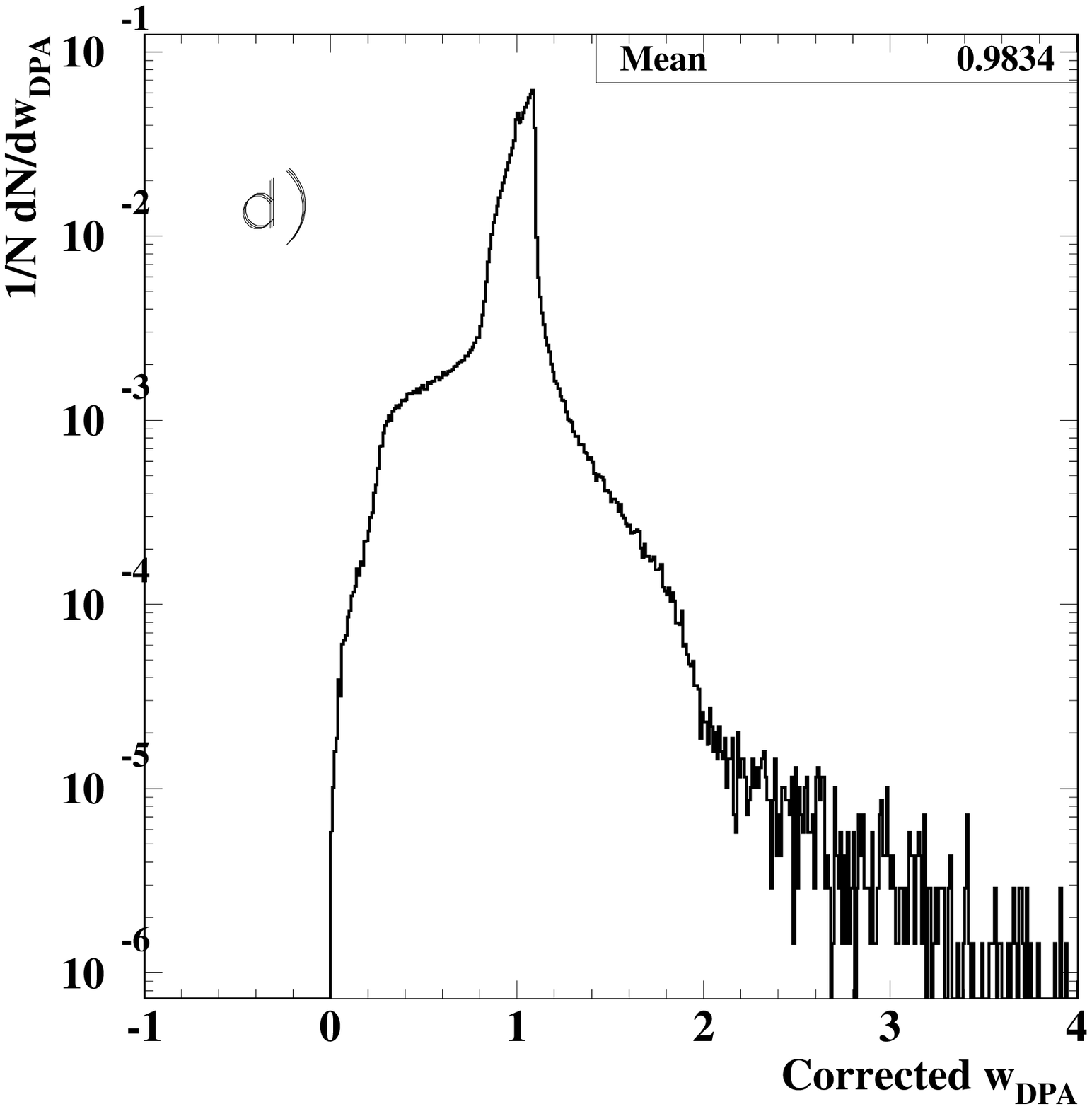}}
  \end{array}
  \)
\end{center}
\caption{Distribution of weights $w$ derived from a 4-$f$ sample of
  events with a
  $WW$ double resonant contribution generated with {\tt WPHACT} at
  $\sqrt{s} = 189$~GeV. a): Relative weight of the DPA squared matrix
  element to the CC03 one with K-C correction applied; b): Relative
  weight of the CC03 squared matrix element with and without K-C
  correction; c): the global DPA weight; Inset: the region around the
  enhancement at 1, due to the non-CC03 contribution; d): the global
  DPA weight after correction to eliminate negative weights. See text
  for details of the evaluation of the weights shown.}
\label{fig1}
\end{figure}

Figure~\ref{fig1} shows the distributions of these matrix element
ratios for events with a $WW$ double resonant (i.e. CC03) contribution
generated with {\tt WPHACT} and reweighted with {\tt YFSWW}. In plot
a) the DPA weight is shown relative to the CC03 weight with the
Khoze-Chapovsky screened Coulomb ansatz.  The effect of this ansatz is
shown in plot b).  The lower figures, c) and d), show the derived DPA
weight to 4-$f$ as defined in equation~(\ref{weight}); the average value
of the distribution gives the total cross-section reduction induced by
DPA.  The small enhancement at $w_{DPA}=1$ represents the contribution
of the events dominated by the non-CC03 part for which
$|\mbox{CC03}|^2/|4f|^2\approx 0$, essentially CC20 events in the single $W$
region or mixed events where the neutral current component is
dominant.

From plot c) of figure~\ref{fig1}, based on the above definition of
$|4f|^2_{DPA}$, it is evident that there are rare events for which the
total DPA weight is negative. These have been investigated and found
to be events compatible with CC03 kinematics, with a sizeable and
negative interference between the CC03 and non-CC03 diagrams (for the
CC20 class they are about 0.1\% of the total), and whose CC03 part is
very much suppressed by DPA.  Thus, from expression~(\ref{weight}) it
is clear that the global weight can become negative, since the DPA
reweighting does not affect the interference term. The solution we have adopted
to this problem is to replace the $Int.$ term with
$Int. \times \sqrt{|\mbox{CC03}_{DPA}|^2/|\mbox{CC03}|^2}$. This corrects only the
modulus of the interference term and not its phase and it removes in
practice the negative weights. The effect of this correction is shown
in plot d).

While in {\tt YFSWW} the radiation from the $W$s is also generated,
this is not present in normal 4-$f$ generators like {\tt KORALW} or
{\tt WPHACT}. Thus the reweighting cannot account for this effect, but
it considers only the ISR part of the radiation. However, the {\tt
  YFSWW} group has shown in~\cite{KandY} that the reweighting
procedure of 4-$f$ {\tt KORALW} events in the additive scheme
correctly reproduces the main differential distributions of CC03
events obtained with {\tt YFSWW} alone, since at LEP2 energies the
effect of radiation off $W$s is marginal. The robustness of the
approach is thus confirmed.

\section{Initial state radiation issues}
\label{ISR}

In relation~(\ref{weight}) the CC03 matrix element is evaluated as
given by the IBA, and therefore already accounts for part of the
radiation. In the ratio $|\mbox{CC03}|^2/|4f|^2$, the standard ISR factorizes
and, therefore, once the value of $s'$ (the effective squared
centre-of-mass energy of the event with the ISR photons removed) is
fixed for the event, it is not crucial which radiator function is used
for the generation of the detailed kinematics of the photons.  But
when the DPA reweighting is applied, it is better to use the same
function in order to get the same $\mathcal{O}(\alpha)$ results as
implemented in the {\tt YFSWW} calculation, since it is based on
corrections applied to a specific radiator.

For this reason the YFS exponentiation for the ISR in the 4-$f$
generation is used in the phase space regions where the DPA
correction has to be applied. This is achieved by interfacing its
implementation in {\tt KORALW} with the {\tt WPHACT} 4-$f$ generator.
The interface was technically possible thanks to the modularity of the
calculation using the leading logarithm approximation; this allowed
the ISR generation to be isolated from the remaining parts of the
code. The YFS radiator is used only for $s$-channel dominated
processes, whereas {\tt QEDPS} is maintained for the $t$-channel
dominated ones, exploiting in this way the dedicated treatment of the
ISR for these channels in the original {\tt WPHACT} code, where the
$t$ scale (defined by the square of the lowest $t$-channel 
$\gamma$ momentum transfer), is used in this case.

\begin{figure}[hbtp]
\begin{center}
  \scalebox{0.9}{\includegraphics{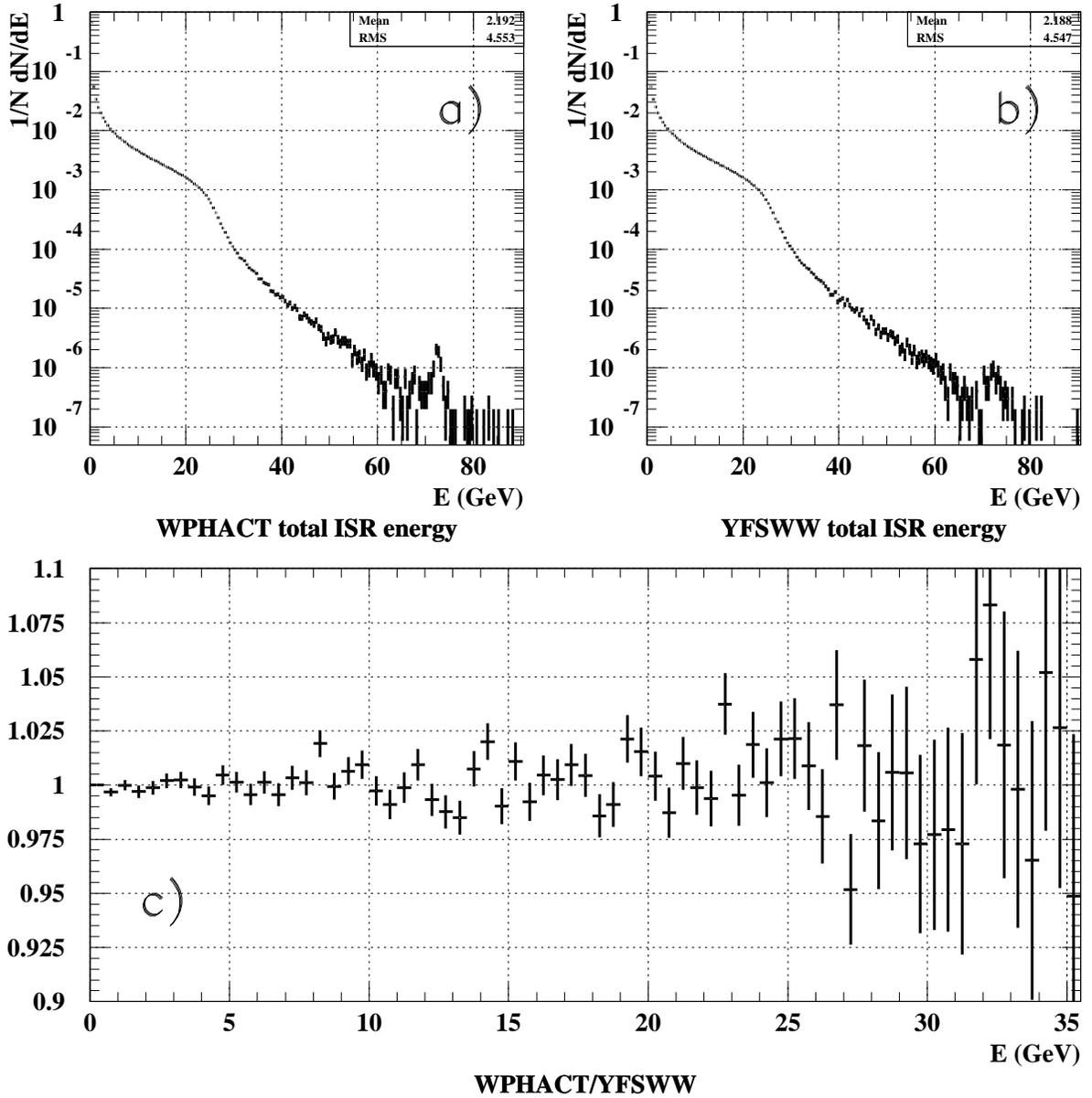}}
\end{center}
\caption{Total ISR photon energy spectrum at Born + ISR level for the process
  $u\bar{d}\mu\bar{\nu}$ evaluated at $\sqrt{s} = 189$~GeV with {\tt
    WPHACT} with YFS exponentiation (a) and with {\tt YFSWW} (b). The
  difference in their mean values is $\Delta(<E_{ISR}>) = 4 \pm 2$
  MeV, where the error is statistical. Plot c) shows the ratio of
  the two spectra.}
\label{figisr1}
\end{figure}

\begin{figure}[hbtp]
  \begin{center}
    \scalebox{0.9}{\includegraphics{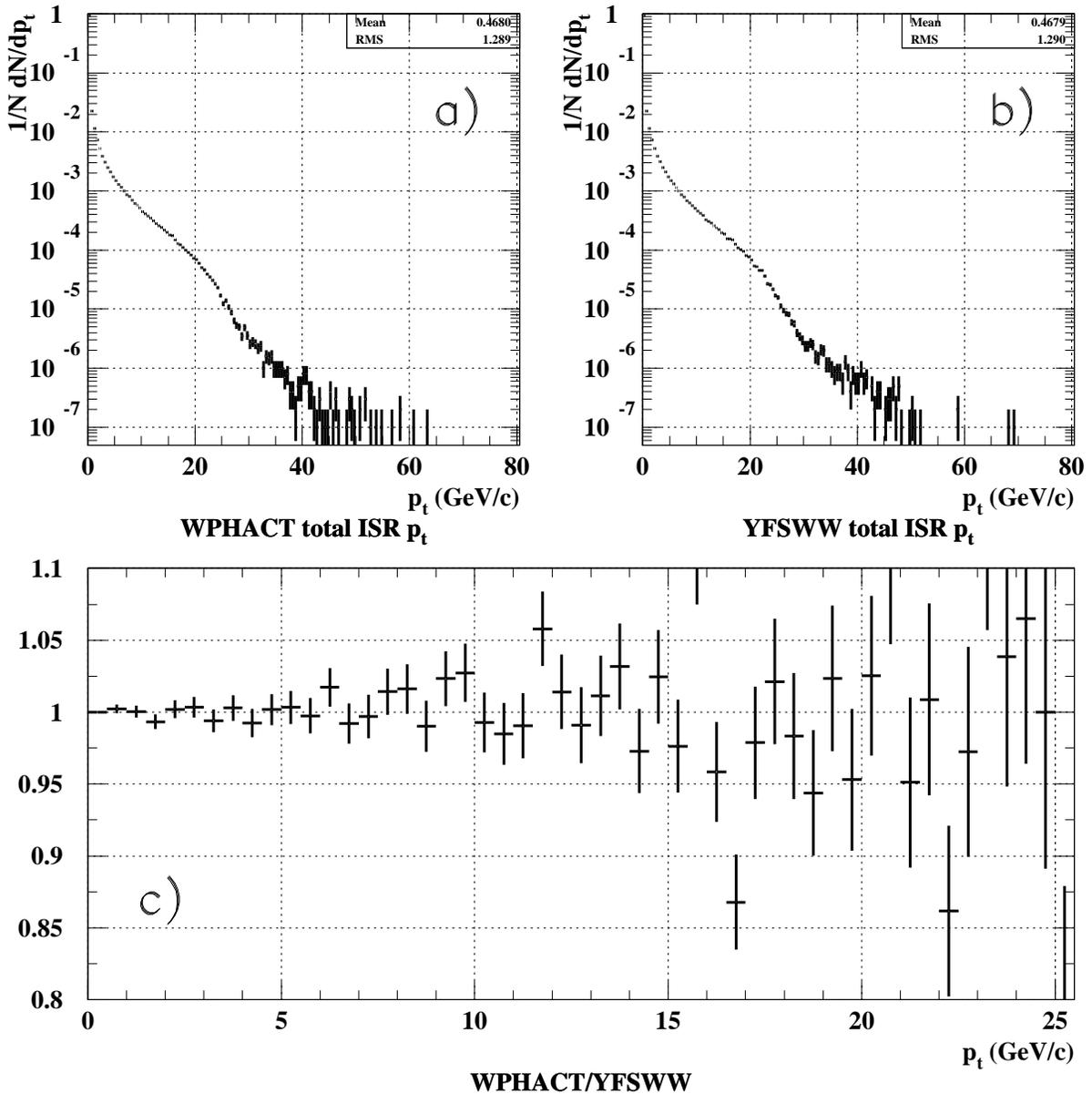}}
\end{center}
\caption{Total photon transverse momentum spectrum at Born + ISR level for the
  process $u\bar{d}\mu\bar{\nu}$ evaluated at $\sqrt{s} = 189$~GeV
  with {\tt WPHACT} with YFS exponentiation (a) and with {\tt YFSWW} (b). The
  difference in mean values is below the statistical uncertainty of 0.6~MeV. 
  Plot c) shows the ratio of the two spectra.}
\label{figisr2}
\end{figure}

\begin{figure}[hbtp]
\begin{center}
  \scalebox{0.9}{\includegraphics{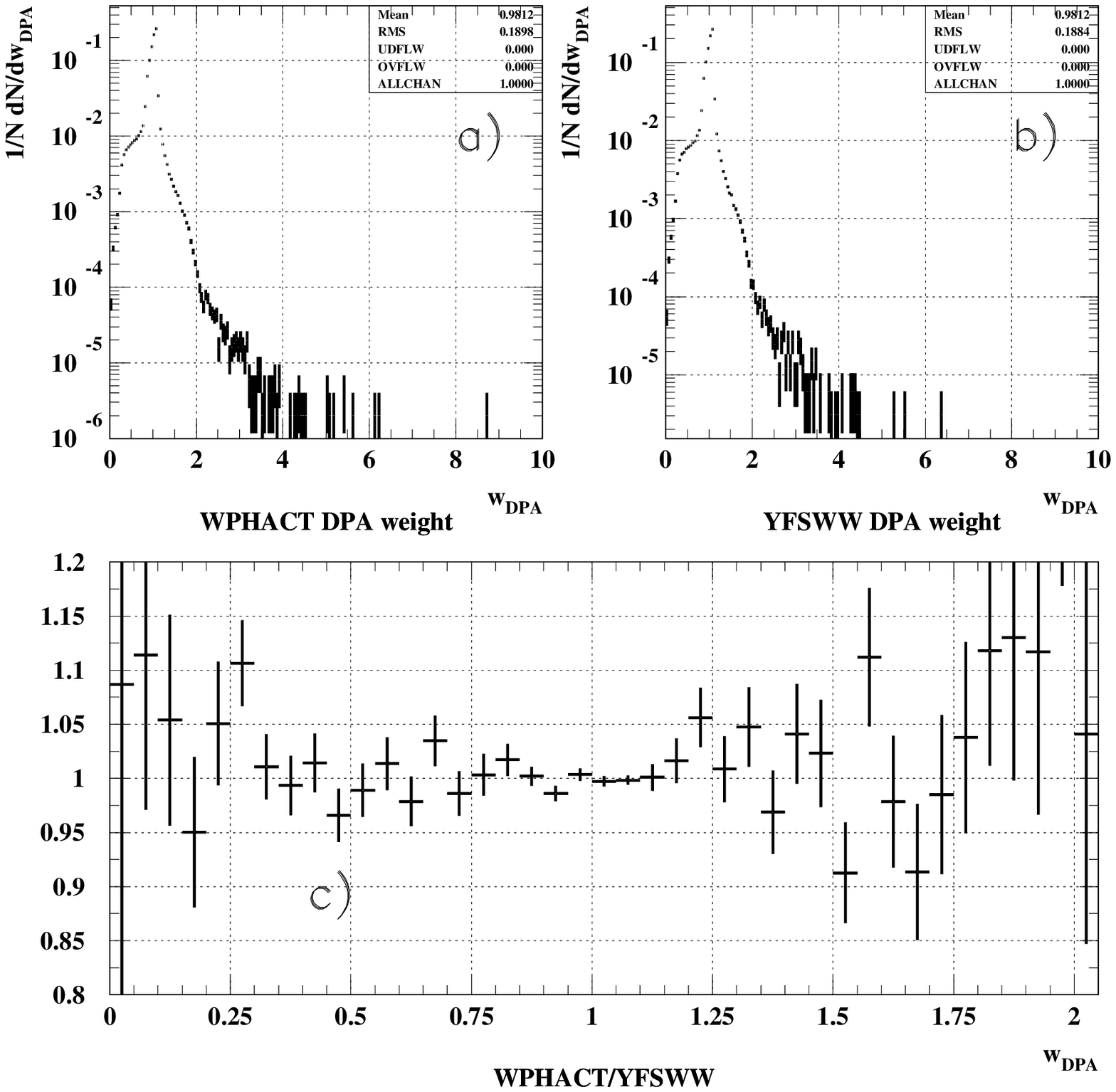}}
\caption{DPA weight $\frac{|\mbox{CC03}_{DPA}|^2}{|\mbox{CC03}|^2}$ distribution for
  the process $u\bar{d}\mu\bar{\nu}$
  evaluated at $\sqrt{s} = 189$~GeV with {\tt WPHACT} with YFS
  exponentiation (a) and with {\tt YFSWW} (b). The
  two distributions are both normalized to unity. The ratio is shown
  in plot c).}
\end{center}
\label{figdpa}
\end{figure}

Extensive tests have shown that the interface of {\tt KORALW}'s
YFS exponentiated ISR in {\tt WPHACT} is consistent with the original
one in {\tt KORALW} and {\tt YFSWW}, both for the total cross-section
and differential distributions. This agreement has also been possible
thanks to very good matching of results between these codes at Born
level for CC03 processes.  Table~\ref{tabisr} compares the total
cross-sections obtained by {\tt WPHACT} with YFS and {\tt YFSWW} in
different configurations for the process $u\bar{d}\mu\bar{\nu}$ at
$\sqrt{s} = 189$~GeV.  Figures~\ref{figisr1} and~\ref{figisr2} show
the comparison of the two generators on the basic ISR photonic
distributions, and figure~\ref{figdpa} shows the comparison of the DPA
weight $\frac{|\mbox{CC03}_{DPA}|^2}{|\mbox{CC03}|^2}$ when computed with the
original {\tt YFSWW} standalone code and with the {\tt WPHACT}/{\tt
YFSWW} tandem.

\begin{table}[h]
\begin{center}
\begin{tabular}{|c|c|c|c|} 
\hline
 & $\sigma(\mbox{{\tt WPHACT}})$ (pb) & $\sigma(\mbox{{\tt YFSWW}})$ (pb) &
 $\sigma(\mbox{{\tt WPHACT}})$/$\sigma(\mbox{{\tt YFSWW}})$ \\
\hline
IBA no CC & 0.59416(7) & 0.59412(7) & 1.0001(2) \\
IBA KC-CS & 0.60817(7) & 0.60730(7) & 1.0014(2) \\
DPA       & 0.59701(8) & 0.59650(8) & 1.0008(2) \\
\hline
\end{tabular}
\caption{Total cross-sections for the process $u\bar{d}\mu\bar{\nu}$
  at $\sqrt{s} = 189$~GeV with {\tt WPHACT} with YFS exponentiation,
  with {\tt YFSWW}, and their ratio. The results for simple IBA, IBA
  with Coulomb corrections in the Khoze-Chapovsky ansatz and in DPA
  are presented. The error from the integration is shown in parenthesis.}
\label{tabisr}
\end{center}
\end{table}

\section{Final state issues: radiation, hadronization, decays}

The standard {\tt WPHACT 2.0} generator provides an interface to the
{\tt PYTHIA} hadronization library via a routine which is based on the
original {\tt PY4FRM} {\tt PYTHIA} interface for 4-$f$ generators.
Moreover, it always fills the standard {\tt HEPEVT} event history
common block~\cite{qcd96}.

The event is transformed to the DELPHI reference frame (the incoming
electron beam goes in the positive $z$ axis direction). The interface
has then been extended in several ways.

\subsection{QED final state radiation}

The QED final state radiation from leptons is treated as in {\tt
  KORALW}/{\tt YFSWW}. When dealing with charged leptons, the {\tt
  PHOTOS} library is called, allowing the production of photons
according to a $\mathcal{O}$$(\alpha^2)$ leading log
calculation.
QED radiation from quarks is treated in the hadronization phase by the
corresponding libraries.  For leptonic $\gamma\gamma$-dominated final
states the QED final state radiation is switched off in {\tt WPHACT},
since it is not appropriate to give a realistic description of the
data in this region.

\subsection{$\tau$ decay}

The $\tau$ decays are described with the {\tt TAUOLA} package, which includes
QED radiative corrections for leptonic decays. In the DELPHI interface the
$\tau$ polarization, which is not provided by {\tt WPHACT}, is defined
according to the event topology. The mother boson of each $\tau$ is identified:
if it is a $W$, the polarization is defined by the $\tau$ charge;
alternatively, if there is a $\tau$ pair coming from a $Z/\gamma^*$, opposite
polarizations are assigned randomly to the taus. {\tt TAUOLA} is used both for
primary taus coming from the hard 4-$f$ process and for those produced in the
hadronization cascade.

\subsection{Quark hadronization}

The default hadronization description is given by {\tt PYTHIA 6.156}
tuned by DELPHI to describe the LEP data at the $Z$ peak~\cite{pytuning}. This
version contains the improvements which
emerged from the LEP2 generator workshop concerning mass effects in
the parton shower and the gluon splitting rate.

Alternatively, the DELPHI tuned versions of {\tt ARIADNE 4.08} and {\tt
  HERWIG 6.2} have also been interfaced, the former using the {\tt
  AR4FRM} interface provided in the library, the latter with a
code based on a former example given by the
authors of {\tt HERWIG} and described in~\cite{qcd96}.  The
availability of different fragmentation models on top of the same
electroweak calculation is an essential feature when studying
systematic effects. The interface allows comparison of the different
colour reconnection and Bose-Einstein correlation schemes implemented
in the above hadronization libraries, -- a topic of great importance
for $WW$ physics.

A problem to be solved when interfacing the 4-$f$ generator with a
hadronization library is the definition of the quark masses. In the
electroweak calculation, at least away from the
$\gamma\gamma$-dominated region, the current algebra masses for the
light quarks are the most suited, allowing a realistic description of
the $\gamma^* \rightarrow q\bar{q}$ decay down to 2 pion masses.  The
heavy quark masses affect the output of the hadronization process, and
their values are chosen in order to get the correct gluon splitting
rate.

In the hadronization phase the definition has to be consistent with
that used in the electroweak calculation, at least from a kinematic
point of view. In {\tt PYTHIA 6.156} (and {\tt ARIADNE}, which is
essentially a different treatment of the gluon radiation part inserted
in {\tt PYTHIA}) the constituent masses for light quarks are provided
as default, but this is not a customary choice for $e^+e^-$ collider
physics, and can be modified. The following set of masses has
therefore been defined to be used both in the electroweak calculations
and in the hadronization phase:

\begin{eqnarray*}
m(\mbox{u}) & = & 0.005 \quad \mbox{GeV}/\mbox{c}^2 \\
m(\mbox{d}) & = & 0.010 \quad \mbox{GeV}/\mbox{c}^2 \\
m(\mbox{s}) & = & 0.200 \quad \mbox{GeV}/\mbox{c}^2 \\
m(\mbox{c}) & = & 1.300 \quad \mbox{GeV}/\mbox{c}^2 \\
m(\mbox{b}) & = & 4.800 \quad \mbox{GeV}/\mbox{c}^2 \, .
\end{eqnarray*}

On the other hand, in {\tt HERWIG} the constituent masses for the
light quarks are used, and since the mass values are tightly linked to
the cluster hadronization model itself, the tuning also depends on
them. In order to avoid inconsistencies in the electroweak part while
using different hadronization models, the above quark masses have been
used in {\tt WPHACT}, and in the interface with {\tt HERWIG} they are
modified into the {\tt HERWIG} ones by imposing 4-momentum
conservation pair by pair according to the colour connection: in this
way the invariant masses of the underlying bosons are preserved, while
in general the quark directions are not.

In order to give a consistent picture of the hadronization throughout
all the processes, the DELPHI tuned {\tt PYTHIA 6.156} with the above
choice of masses has also been used in the 2-$f$ sector (with the {\tt
  KK2f}~\cite{kk2f} generator) and in the Higgs sector (with the {\tt
  HZHA}~\cite{hzha} generator).

As mentioned before, the above choice is used in all the phase space
regions not dominated by multiperipheral diagrams.  In the latter case
the constituent masses allow a better description of reality, acting
as an effective cut-off on the cross-section. Therefore the following
values are used for the light quark masses:

\begin{eqnarray*}
m(\mbox{u}) & = & 0.3 \quad \mbox{GeV}/\mbox{c}^2 \\
m(\mbox{d}) & = & 0.3 \quad \mbox{GeV}/\mbox{c}^2 \\
m(\mbox{s}) & = & 0.5 \quad \mbox{GeV}/\mbox{c}^2 \, .
\end{eqnarray*}

The colour connection scheme used in the presence of mixed charged
current and neutral current 4-quark final states exploits the separate
generation of the charged and neutral currents in {\tt WPHACT} in such
a way as to generate the correct proportion of events of each kind,
including the interference between the two: in this way it is known
{\it a priori} whether the current event has to be hadronized, for
instance, as $u\bar{d}d\bar{u}$ (i.e.  $WW$) or as $u\bar{u}d\bar{d}$
(i.e.  $ZZ$).

The only ambiguous case is in the presence of 4 identical quarks $q_1
\bar{q}_1 q_2 \bar{q}_2 (q_1=q_2)$, where the approach adopted is as
discussed in~\cite{qcd96} and also used in {\tt PYTHIA}: the quark
pairing is randomly chosen according to the relative probability of
each of the two configurations as given by the ratio of the squared
amplitudes $|\mathcal{M}$$(q_1\bar{q}_1
+q_2\bar{q}_2)|^2/|\mathcal{M}$$(tot)|^2$ and
$|\mathcal{M}$$(q_1\bar{q}_2+q_2\bar{q}_1)|^2/|\mathcal{M}$$(tot)|^2$.
The event pairing choice is recorded in the standard output of the
generator.

\subsection{Low $q\bar{q}$ mass system hadronization}

The string model is not suitable to describe the hadronization of $q\bar{q}$
systems below a mass of 2~GeV/$\mbox{c}^2$. The package in~\cite{Boonekamp} provides a
description of the hadronization from the $\gamma^* \rightarrow q\bar{q}$
process in this low mass region both due to the presence of hadronic
resonances (with subsequent decays described by {\tt PYTHIA}) and in the
continuum, based on experimental $e^+e^-$ data at low energy. The package has
been fully interfaced with {\tt WPHACT} and, when the quark pairing algorithm
described above gives rise to a $q\bar{q}$ pair of mass below 2~GeV/$\mbox{c}^2$, it is
used instead of the other hadronization models to produce the final hadronic
state.

\section{Phase space cuts and matching with $\gamma\gamma$ generators}
\label{par:phsp}

Ideally the 4-$f$ generation should cover the whole experimentally
accessible phase space without any cut. In practice a compromise has
to be found between the needs of the physics analyses and the
numerical and physical reliability of the calculation, both in terms
of matrix element evaluation and of phase space integration. Although
the use of FORTRAN quadruple precision arithmetic can significantly
boost the numerical stability in delicate regions of the phase space
(very low electron angles and very low $\gamma^*$ masses), it is more
CPU consuming and not available in all compilers.  In any case, one
does not want to produce a huge fraction of generally uninteresting
events.

A dedicated study has therefore been performed to define a set of cuts
which can realize the above compromise. The basic set of cuts on
fermion-antifermion invariant masses and fermion (and antifermion)
energies applied to all the final states is shown in
table~\ref{tab:cuts}.

\begin{table}[h]
  \begin{center}
    \begin{tabular}{|c|}
      \hline
      Phase space cuts \\
      \hline \hline
      Fermion-antifermion invariant mass \\
      \hline
      $m(q_i\bar{q_i}) > \mbox{max}( 2\,m_{\pi} , 2\,m_{q_i} )$ \\
      $m(q_i\bar{q_j}) > \mbox{max}( 2 \, \mbox{GeV}/\mbox{c}^2 , m_{q_i}+m_{q_j})$ \\
      $m(e^+e^-) > 0.2 \, \mbox{GeV}/\mbox{c}^2$ \\
      \hline \hline
      Fermion energy \\
      \hline
      $E(q) > 1 \, \mbox{GeV}$ \\
      $E(e) > 1 \, \mbox{GeV}$ if $5^\circ < \theta_e < 175^\circ$ \\
      $E(\mu) > 1 \, \mbox{GeV}$ if $2^\circ < \theta_{\mu} < 178^\circ$ \\
      $E(\tau) > 1 \, \mbox{GeV}$ if $2^\circ < \theta_{\tau} < 178^\circ$ \\
      \hline
    \end{tabular}
    \caption{Phase space cuts common to all processes. $q$ means
      quark, $e$, $\mu$ and $\tau$ the charged leptons. $\theta$ is
      the polar angle of the fermion. Implicit
      invariant mass cuts given by the fermion masses are not listed.}
    \label{tab:cuts}
  \end{center}
\end{table}

In some classes of events additional requirements have been imposed: \\

\noindent CC18 ($e\nu\mu\nu , e\nu\tau\nu$): $5^\circ < \theta_e <
175^\circ$ or $5^\circ < \theta_{\mu,\tau} < 175^\circ$, i.e. there
must be at least one visible lepton; \newline
MIX56 ($e\nu e\nu$): $5^\circ < \theta_{e^-} < 175^\circ$ or 
   $5^\circ <\theta_{e^+} < 175^\circ$. \\

As explained in the previous paragraphs, particular care is needed
when treating the NC48 ($eeqq, ee\mu\mu, ee\tau\tau$) and NC144
($eeee$) final states, which get contributions from the
multiperipheral diagrams, and in particular by the direct photon
component of the $\gamma\gamma$ process.

In the description of the $eeqq$ final state, most general 4-$f$
generators correctly compute the unresolved photon part of the
$\gamma\gamma$ interaction, dominant at high photon virtualities,
where the photon is treated as a pointlike particle. However, they
usually cannot properly handle the resolved photon part, i.e. the part
where the hadronic content of the photon becomes relevant, and is
described by partonic distributions or by the vector meson dominance
ansatz, depending on the virtuality range.  Dedicated codes model the
process in a more realistic way in the phase space region where the
resolved photon component starts to be important; usually the
non-$\gamma\gamma$ related Feynman diagrams are neglected, since in
this region they are highly suppressed. In DELPHI, this part of the
phase space has been treated using {\tt PYTHIA 6.143}.  The pure
direct photon part, corresponding to the multiperipheral diagrams with
two photons exchanged in the $t$-channel, and the diagram involving at
least one resolved photon are added incoherently. This splitting
allows use of the full 4-$f$ calculation for the former in the phase
space regions where the non-multiperipheral contribution is still
relevant, preserving at the same time {\tt PYTHIA}'s description of
the latter everywhere.

Three different regions of the phase space have been identified for
$eeff$ final states:

\begin{itemize}
\item the ``4-$f$ like'' region, where the multiperipheral
  contribution is not dominant and {\tt WPHACT} with the current
  algebra masses for the light quarks can be considered fully
  reliable;
\item the ``$\gamma\gamma$ like'' region, where the multiperipheral
  contribution starts to be the dominant one but other electroweak
  contributions are important and need to be properly described; in
  this region {\tt WPHACT} with constituent masses for light quarks is
  used, as discussed above, and the QED final state radiation from
  leptons generated with {\tt PHOTOS} is switched off;
\item the pure $\gamma\gamma$ region, described with the dedicated
  codes.  For the fully leptonic final states, {\tt
    BDKRC}~\cite{bdkrc} has been used for the $ee\mu\mu$ and
  $ee\tau\tau$ final states, since it contains the matrix element with
  $\mathcal{O}(\alpha)$ radiative corrections.  For the $eeee$ final
  state the dedicated code {\tt BDK}~\cite{bdk} is used, mainly for
  reasons of technical reliability.
\end{itemize}

The detailed description of the phase space cuts defining these three
regions is given in appendix~\ref{app1}.

\begin{figure}[hbtp]
\begin{center}
  \scalebox{0.9}{\includegraphics{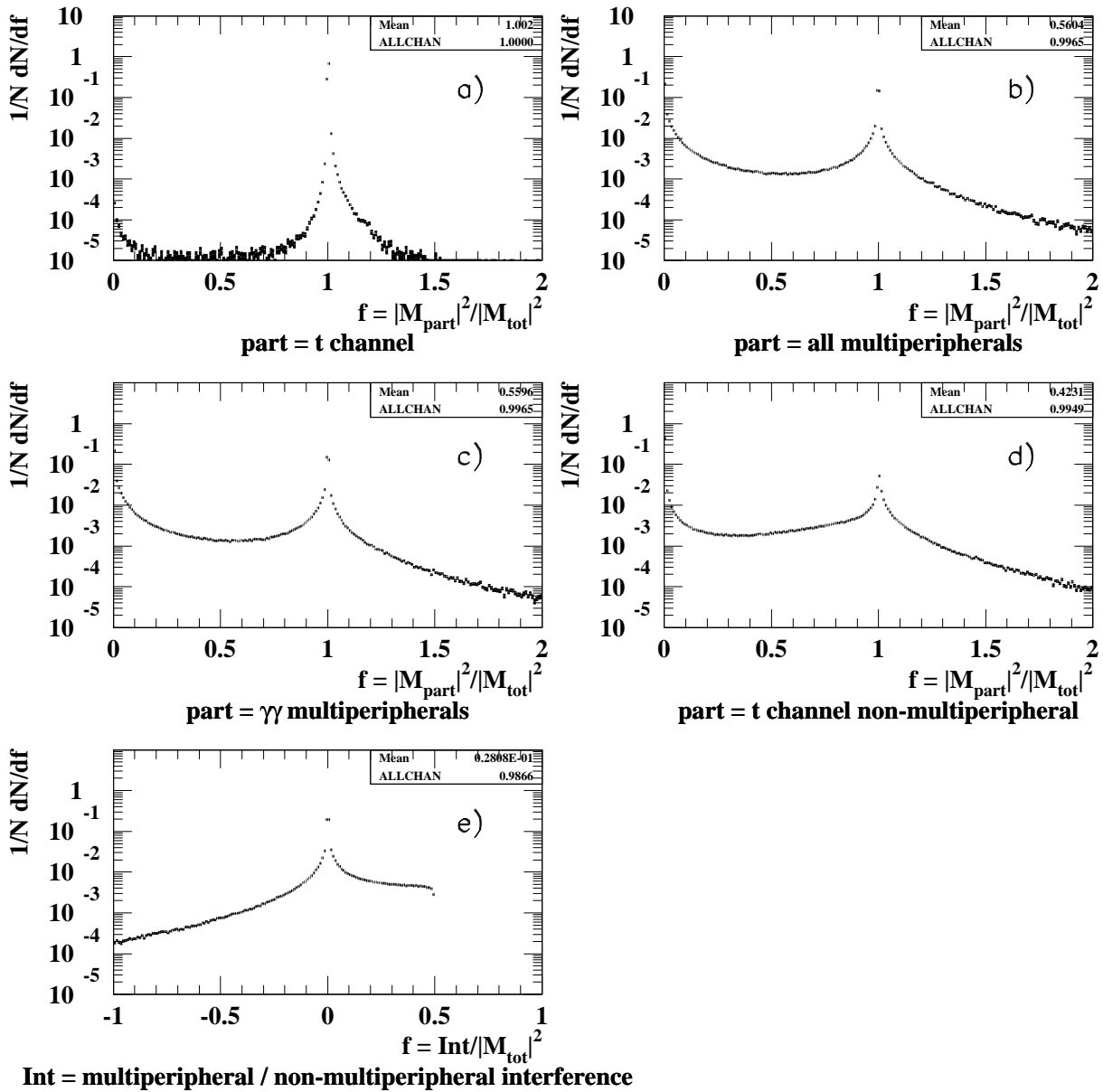}}
\caption{a)--d): Spectra of the fractions $f =
  |\mathcal{M}$$_{part}|^2/|\mathcal{M}$$_{tot}|^2$ of the total squared
  event matrix element $|\mathcal{M}$$_{tot}|^2$ corresponding to
  different subsets of Feynman diagrams (labelled as $part$) for events
  belonging to all the $\gamma\gamma$ compatible final states (i.e.
  $eeqq$ and $eell$) in the ``$\gamma\gamma$ like" region at $\sqrt{s}
  = 189$~GeV: a) $t$-channel diagrams, b) multiperipheral diagrams, c)
  multiperipheral diagrams with only photon exchange (i.e. direct
  photon $\gamma\gamma$ ), d) $t$-channel non-multiperipheral diagrams;
  e) interference between the multiperipheral part and the
  remaining amplitude.}
\label{fig:ggme}
\end{center}
\end{figure}

Figure~\ref{fig:ggme} shows, for events belonging to all the
$\gamma\gamma$ compatible final states (i.e.  $eeqq$ and $eell$) at
$\sqrt{s} = 189$~GeV, the fractions of the total squared event matrix
element in the ``$\gamma\gamma$ like'' region corresponding to
different subsets of Feynman diagrams. It can be seen
(figure~\ref{fig:ggme}a) that the $t$-channel component is largely
dominant, as expected. The multiperipheral diagrams constitute a large
fraction of the total (figure~\ref{fig:ggme}b) -- almost half of the
events' matrix elements are almost identical to the purely
multiperipheral ones; -- furthermore those corresponding to genuine
direct photon $\gamma\gamma$ diagrams are clearly the dominant part of
the multiperipherals (figure~\ref{fig:ggme}c). But it can also be seen
that there is a sizeable component where the non-multiperipheral
$t$-channel diagrams dominate (figure~\ref{fig:ggme}d), and the
interference between these and all the rest is not negligible in about
half of the events (figure~\ref{fig:ggme}e). Since the $s$-channel
hardly contributes in this region, this interference is essentially
between the multiperipheral and non-multiperipheral $t$-channel
amplitudes.

One must be aware that the cuts used in the ``$\gamma\gamma$ like"
region still leave some numerical instability of the order of a few
per cent when using double precision. This is found comparing, for
instance, {\tt WPHACT} double and quadruple precision results for $e e
u\bar u$ single tag events with $10^\circ < \theta_e < 170^\circ$ and
$3<m(u\bar u)<40$~GeV/$\mbox{c}^2$ (see appendix~\ref{app1}) at various energies.
Normally the difference is of the order of 2\%, but values as high as
6\% have been found. This difference diminishes and rapidly disappears
if the lower value of m($u\bar u$) is increased by a few GeV/$\mbox{c}^2$.
The above numerical instability is in any case well below the
estimated theoretical uncertainty. A simple example of this is given
by the fact that, for $eeu\bar{u}$, the difference in using
constituent or current quark masses already amounts to 14\%.

The cross-sections obtained with the above phase space coverage at a
typical LEP2 energy, $\sqrt{s} = 199.5$~GeV, are given in
appendix~\ref{app2} together with a detailed description of the
standard input parameter set used.

\section{Matrix elements for reweighting}
\label{merew}

It is common practice to define cross-sections for 4-$f$ processes considering
only subsets of Feynman diagrams: CC03, NC02, $t$-channel single $W$.  Although
unphysical, as all the diagrams take part in the event probability, these
definitions reflect the dominant component in the phase space region under
study. Thus correction factors have to be evaluated to extract the
relevant part of the experimentally measured cross-section. It is therefore
useful to compute, for each event, the squared matrix elements
$|\mathcal{M}$$|^2$ corresponding to several potentially interesting subsets of
Feynman diagrams. They can be used to determine their respective components
in the unweighted sample. Moreover, calculating  for each subset the matrix
elements with the remaining graphs allows the interference between the two to
be evaluated, thus completing the information.

The squared matrix elements corresponding to the following subsets are
given in the standard DELPHI output:

\begin{enumerate}
\item CC03 (i.e. $WW$),
\item $t$-channel component,
\item NC02 (i.e. $ZZ$),
\item $Z\gamma^*$,
\item $\gamma^*\gamma^*$,
\item NC08 (i.e. $ZZ + Z\gamma^* + \gamma^*\gamma^*$),
\item multiperipheral (neutral currents only),
\item multiperipheral $\gamma\gamma$ only,
\item $t$-channel non-multiperipheral component.
\end{enumerate}

In CC03-related analyses it is interesting to know the behaviour of
the squared matrix element as a function of several physical input
parameters: $W$ mass, $W$ width, trilinear gauge coupling (TGC) parameters. The
$W$ width in the Standard Model is not an independent parameter, but is defined
by the $W$ mass and couplings. Nevertheless, in the experimental fits, it can
be useful to treat it as a free parameter: in {\tt WPHACT} this possibility is
provided, keeping the $W$ mass and the couplings at their Standard Model
values.

A compact and precise way of describing the functional dependence of
the squared matrix element on the $W$ mass and width is adopted.  A
suitable number of coefficients of the Chebyshev polynomial expansion
of $\log_{10}|\mathcal{M}$$|^2$ is computed in the physically
interesting interval: [78.5, 82.5]~GeV/$\mbox{c}^2$ for the mass, and 
[0.2, 6.2]~GeV/$\mbox{c}^2$ for the width.


The functional dependence of the squared matrix element on the
TGCs (the parametrization $\Delta g^Z_1$,
$\Delta\kappa_{\gamma}$, $\lambda_{\gamma}$ in~\cite{tgcpar} is used)
is also given as part of the standard output. Since the cross-section
is a quadratic function in term of these TGC parameters, the
coefficients of this quadratic parametrization are computed by solving
a linear system of equations obtained by evaluating the cross-section
for a suitable number of combinations of TGC values ({\it e.g}, 10
coefficients to describe the variation of 3 parameters).

\section{Summary}
We have described the setup for 4-$f$ event generation chosen by the
DELPHI collaboration for LEP2 measurements. The main difficulty in
its construction consisted in interfacing in the optimal way the best
features which were available at the end of the LEP2 generator
workshop for the various 4-$f$ processes, in order to describe every corner of
the phase space as accurately as possible.  This is very important for
physics measurements, with particular attention to the $W$ sector, but
also for the best evaluation of backgrounds to search processes.

\section{Acknowledgements}

We are grateful to S.~Jadach, W.~Placzek, M.~Skrzypek, B.~F.~L.~Ward
and Z.~Was for useful discussions and for the code used to implement
the YFS exponentiated ISR in the DELPHI version of {\tt WPHACT}.  We
thank M.~Boonekamp and S.~Todorova for discussions and
collaboration on the phase space cuts and $\gamma\gamma$ matching
definition, E.~Accomando and E.~Maina for their
collaboration and advice, and R.~Sekulin for his careful reading and
his comments on the manuscript.

\newpage

\appendix
\section{Appendix: phase space cuts for 4-$f$ - $\gamma\gamma$
  matching}
\label{app1}

The ``4-$f$ like'' region is defined as:
  \begin{itemize}
  \item $eeff$ where $f \neq e$ double tag, i.e. both electron and
    positron have polar angle\footnote
    {Here and below, equivalent conditions are implied on the
    supplements of the quoted values of the polar angle, i.e. 
    $\theta > x \Rightarrow \theta < 180^\circ-x$ and $\theta < x
    \Rightarrow \theta > 180^\circ-x$.}
    of $\theta_e > 10^\circ$;
  \item $eeff$ where $f \neq e$ single tag, where the visible electron has a
    polar angle of $\theta_e > 10^\circ$ and $m(ff) > 40 \,
    \mbox{GeV}/\mbox{c}^2$;
  \item $eeee$ with 3 or 4 visible electrons, i.e. with a polar angle
    of $\theta_e > 10^\circ$.
  \end{itemize}

The ``$\gamma\gamma$ like'' region is defined as:
  \begin{itemize}
  \item $eeqq$ single tag, where the visible electron has a polar angle of
    $\theta_e > 10^\circ$ and $3 < m(qq) < 40 \, \mbox{GeV}/\mbox{c}^2$;
  \item $ee\mu\mu$ single tag, where the visible electron has a polar
    angle of $\theta_e > 10^\circ$, $m(\mu\mu) < 40 \, \mbox{GeV}/\mbox{c}^2$
    and at least one muon has $\theta_{\mu} > 2^\circ$;
  \item $ee\tau\tau$ single tag, where the visible electron has a
    polar angle of $\theta_e > 10^\circ$, $m(\tau\tau) < 40 \,
    \mbox{GeV}/\mbox{c}^2$ and at least one tau has $\theta_{\tau} > 2^\circ$;
  \item $eeff$ where $f \neq e$ double tag, i.e. both electron and positron
    have polar angle of $2^\circ < \theta_e < 10^\circ$;
  \item $eeff$ where $f \neq e$ single tag, i.e. the visible electron
    has a polar angle of $2^\circ < \theta_e < 10^\circ$
    and $m(ff) > 40 \, \mbox{GeV}/\mbox{c}^2$;
  \item $eeff$ where $f \neq e$ no tag, with both electron and
    positron with a polar angle of $\theta_e < 2^\circ$ and
    $m(ff) > 40 \, \mbox{GeV}/\mbox{c}^2$;
  \item $eeee$ quadruple tag, where the tagged electrons have a polar angle of
    $\theta_e > 2^\circ$ but not all of them have $\theta_e > 10^\circ$;
  \item $eeee$ triple tag, where the tagged electrons have $\theta_e > 2^\circ$
    but not all of them have $\theta_e > 10^\circ$, and the minimum of the
    invariant masses $m(e^+e^-)$ is above 40~GeV/$\mbox{c}^2$;
  \item $eeee$ double tag, where the tagged electrons have $\theta_e >
    2^\circ$ and the minimum of the invariant masses $m(e^+e^-)$ is above
    40~GeV/$\mbox{c}^2$.
  \end{itemize}

The pure $\gamma\gamma$ region is the one complementary to the two regions
defined above.

\newpage

\section{Appendix: Input parameter set}
\label{app2}

For the standard event generation the following input parameters have
been used:

\begin{eqnarray*}
  G_{\mu} & = & 1.16639 \times 10^{-5} \, \mbox{GeV}^{-2} \\
  M_{W} & = & 80.4 \, \mbox{GeV}/\mbox{c}^2 \\
  M_{Z} & = & 91.187 \, \mbox{GeV}/\mbox{c}^2 \\
  m_{t} & = & 175 \, \mbox{GeV}/\mbox{c}^2 \\
  M_{H} & = & 115 \, \mbox{GeV}/\mbox{c}^2 \, \, \footnotemark \\
  \alpha(Q^2 = 0) & = & 1/137.0359895 
\end{eqnarray*}

\footnotetext{Used only in the DPA calculation; at Born level no
  diagram with Higgs boson exchange is considered.}

The $G_{\mu}$ renormalization scheme is used, the running of $\alpha$
is used in the low invariant mass regions and the running width is
used for the boson propagators. The lepton masses are fixed at their
PDG values~\cite{PDG}. The naive QCD correction is used for the boson
width correction.

As an example of the results obtained with the above parameters and
the phase space cuts discussed in the text, the cross-sections for all
the processes at $\sqrt{s} = 199.5$~GeV are listed in
tables~\ref{tab:ccxsec}, \ref{tab:ncxsec} and \ref{tab:ggxsec}. The
total generated cross-section is $82.95 \pm 0.03$ pb, where the error is 
purely statistical, corresponding to the accuracy of the integration.

\begin{table}[h]
  \begin{center}
    \begin{tabular}{|c|l l l l|}
      \hline
      \multicolumn{5}{|c|}{} \\
      \multicolumn{5}{|c|}{\bf CC}\\
      \multicolumn{5}{|c|}{} \\
      \hline \hline
      process type & final state & cross-section (pb) & final state &
      cross-section (pb) \\
      \hline
      \rule{0cm}{0.5cm} 
       CC09 & $\mu^-$ $\bar\nu_\mu$ $\nu_\tau$ $\tau^+$ & 0.40756(9)
            &  & \\
       \hline \rule{0cm}{0.5cm} 
       CC18 & $e^-$  $\bar\nu_e$  $\nu_\mu$  $\mu^+$ & 0.50649(9) 
            & $e^-$  $\bar\nu_e$  $\nu_\tau$  $\tau^+$ & 0.50333(9) \\
       \hline \rule{0cm}{0.5cm} 
       CC10 & $\mu^-$  $\bar\nu_\mu$  $u$  $\bar d$ & 1.2096(3) 
            & $\tau^-$  $\bar\nu_\tau$  $u$  $\bar d$ & 1.2087(3) \\
            & $\mu^-$  $\bar\nu_\mu$  $c$  $\bar s$ & 1.2077(3)
            & $\tau^-$  $\bar\nu_\tau$  $c$  $\bar s$ & 1.2058(3) \\
            & $\mu^-$  $\bar\nu_\mu$  $u$  $\bar s$ & 0.06282(1) 
            & $\tau^-$  $\bar\nu_\tau$  $u$  $\bar s$ & 0.06276(1) \\
            & $\mu^-$  $\bar\nu_\mu$  $c$  $\bar d$ & 0.06274(1) 
            & $\tau^-$  $\bar\nu_\tau$  $c$  $\bar d$ & 0.06261(1) \\
            & $\mu^-$  $\bar\nu_\mu$  $c$  $\bar b$ & 0.0020876(5) 
            & $\tau^-$  $\bar\nu_\tau$  $c$  $\bar b$ & 0.0020860(5) \\
       \hline \rule{0cm}{0.5cm} 
       CC20 & $e^-$  $\bar\nu_e$  $u$  $\bar d$ & 1.5300(3)
            & $e^-$  $\bar\nu_e$  $c$  $\bar d$ & 0.07786(1) \\
            & $e^-$  $\bar\nu_e$  $c$  $\bar s$ & 1.4981(3)       
            & $e^-$  $\bar\nu_e$  $c$  $\bar b$ & 0.0025850(6) \\
            & $e^-$  $\bar\nu_e$  $u$  $\bar s$ & 0.07936(1) & & \\
       \hline \rule{0cm}{0.5cm}                                      
       CC11 & $s$  $\bar c$  $u$  $\bar d$ & 3.5744(7)       
            & $d$  $\bar c$  $c$  $\bar b$ & 0.00032050(7) \\           
            & $s$  $\bar c$  $u$  $\bar s$ & 0.18562(5) 
            & $b$  $\bar c$  $u$  $\bar d$ & 0.006187(1) \\         
            & $s$  $\bar c$  $c$  $\bar d$ & 0.18522(5)
            & $b$  $\bar c$  $u$  $\bar s$ & 0.00032124(7) \\           
            & $s$  $\bar c$  $c$  $\bar b$ & 0.006172(1)
            & $s$  $\bar u$  $u$  $\bar s \ ^{\ **}$ & 0.004831(1) \\
            & $d$  $\bar u$  $u$  $\bar s$ & 0.18606(5) 
            & $d$  $\bar c$  $c$  $\bar d \ ^{\ **}$ & 0.004809(1) \\
            & $d$  $\bar c$  $u$  $\bar d$ & 0.18558(5)        
            & $b$  $\bar c$  $c$  $\bar b \ ^{\ **}$ & 0.000005337(2) \\
            & $d$  $\bar c$  $u$  $\bar s$ & 0.009638(1)          
            & & \\
      \hline \hline
      \multicolumn{5}{|c|}{} \\
      \multicolumn{5}{|c|}{\bf Mixed}\\
      \multicolumn{5}{|c|}{} \\
      \hline \hline
      process type & final state & cross-section (pb) & final state &
      cross-section (pb) \\
      \hline \rule{0cm}{0.5cm} 
      MIX19 & $\mu^-$  $\mu^+$  $\nu_\mu$  $\bar\nu_\mu$ & 0.22697(4)
            & $\tau^-$  $\tau^+$  $\nu_\tau$  $\bar\nu_\tau$ & 0.21412(4) \\
      \hline \rule{0cm}{0.5cm} 
      MIX56 & $e^-$  $e^+$  $\nu_e$  $\bar\nu_e$ & 0.4197(1) & & \\
      \hline \rule{0cm}{0.5cm} 
      MIX43 & $d$  $\bar d$  $u$  $\bar u$ & 1.9070(3)
            & $s$  $\bar s$  $c$  $\bar c$ & 1.8582(3) \\
      \hline
    \end{tabular} 
    \caption{Cross-sections for all the charged and mixed current 4-$f$
      processes at $\sqrt{s} = 199.5$~GeV with the phase space cuts
      discussed in the text. For CC processes the charge
      conjugate final state is also included. Those labelled with $^{**}$
      are only the CC contribution to the full process; the neutral
      current part is given in the NC table. The errors quoted are
      purely statistical, corresponding to the accuracy of the integration.}
    \label{tab:ccxsec}
  \end{center}
\end{table}

\begin{table}[hbtp]
  \begin{center}
    \begin{tabular}{|c|l l l l|}
      \hline
      \multicolumn{5}{|c|}{} \\
      \multicolumn{5}{|c|}{\bf NC (not $\gamma\gamma$ compatible)}\\
      \multicolumn{5}{|c|}{} \\
      \hline \hline
      process type & final state & cross-section (pb) & final state &
      cross-section (pb) \\
      \hline \rule{0cm}{0.5cm} 
      NC06 & $\nu_\mu$  $\bar\nu_\mu$  $\nu_\tau$  $\bar\nu_\tau$ & 0.008784(2)
      & & \\
      \hline \rule{0cm}{0.5cm} 
      NC12 & $\nu_\mu$  $\bar\nu_\mu$  $\nu_e$  $\bar\nu_e$ & 0.009374(3)
           & $\nu_\tau$  $\bar\nu_\tau$  $\nu_e$  $\bar\nu_e$ & 0.009378(3) \\
      \hline \rule{0cm}{0.5cm} 
      NC12 & $\nu_\mu$  $\bar\nu_\mu$  $\nu_\mu$  $\bar\nu_\mu$ & 0.004358(1)
           & $\nu_\tau$  $\bar\nu_\tau$  $\nu_\tau$  $\bar\nu_\tau$ &
      0.004358(1) \\
      \hline \rule{0cm}{0.5cm} 
      NC36 & $\nu_e$  $\bar\nu_e$  $\nu_e$  $\bar\nu_e$ & 0.004845(2) &&\\
      \hline \rule{0cm}{0.5cm} 
      NC10 & $u$  $\bar u$  $\nu_\mu$  $\bar\nu_\mu$ & 0.04149(3)
           & $c$  $\bar c$  $\nu_\mu$  $\bar\nu_\mu$ & 0.02611(1) \\  
           & $u$  $\bar u$  $\nu_\tau$  $\bar\nu_\tau$ & 0.04149(3)
           & $c$  $\bar c$  $\nu_\tau$  $\bar\nu_\tau$ & 0.02611(1) \\
      \hline \rule{0cm}{0.5cm} 
      NC19 & $u$  $\bar u$  $\nu_e$  $\bar\nu_e$ & 0.06211(4) 
           & $c$  $\bar c$  $\nu_e$  $\bar\nu_e$ & 0.03291(1) \\
      \hline \rule{0cm}{0.5cm} 
       NC64 & $u$  $\bar u$  $u$  $\bar u$ & 0.0776(1)
            & $c$  $\bar c$  $c$  $\bar c$ & 0.03741(1) \\
       \hline \rule{0cm}{0.5cm} 
      NC32 & $u$  $\bar u$  $c$  $\bar c$ & 0.11257(8) &&\\
      \hline \rule{0cm}{0.5cm} 
      NC10 & $\mu^-$  $\mu^+$  $\nu_\tau$  $\bar\nu_\tau$ & 0.022940(9)
           & $\tau^-$  $\tau^+$  $\nu_\mu$  $\bar\nu_\mu$ & 0.010351(2) \\
      \hline \rule{0cm}{0.5cm} 
      NC20 & $e^-$  $e^+$  $\nu_\mu$  $\bar\nu_\mu$ & 0.11120(6)
           & $e^-$  $e^+$  $\nu_\tau$  $\bar\nu_\tau$ & 0.11120(6) \\
      \hline \rule{0cm}{0.5cm}                                             
      NC19 & $\mu^-$  $\mu^+$  $\nu_e$  $\bar\nu_e$  & 0.03811(2)
           & $\tau^-$  $\tau^+$  $\nu_e$  $\bar\nu_e$ & 0.013485(4) \\  
      \hline \rule{0cm}{0.5cm}                                             
      NC24 & $\mu^-$  $\mu^+$  $u$  $\bar u$ & 0.0775(2)
           & $\tau^-$  $\tau^+$  $u$  $\bar u$ & 0.04264(3) \\         
           & $\mu^-$  $\mu^+$  $c$  $\bar c$  & 0.06007(5)
           & $\tau^-$  $\tau^+$  $c$  $\bar c$ & 0.02891(1) \\         
      \hline \rule{0cm}{0.5cm}                                              
      NC19 & $d$  $\bar d$  $\nu_e$  $\bar\nu_e$ & 0.03291(3)
           & $b$  $\bar b$  $\nu_e$  $\bar\nu_e$ & 0.03058(2) \\       
           & $s$  $\bar s$  $\nu_e$  $\bar\nu_e$ & 0.022758(9)        
        &&\\                                             
      \hline \rule{0cm}{0.5cm}                                              
      NC10 & $d$  $\bar d$  $\nu_\mu$  $\bar\nu_\mu$ & 0.02672(2)
           & $d$  $\bar d$  $\nu_\tau$  $\bar\nu_\tau$ & 0.02672(2) \\ 
           & $s$  $\bar s$  $\nu_\mu$  $\bar\nu_\mu$ & 0.02550(1)    
           & $s$  $\bar s$  $\nu_\tau$  $\bar\nu_\tau$ & 0.02550(1) \\ 
           & $b$  $\bar b$  $\nu_\mu$  $\bar\nu_\mu$ & 0.021165(7)    
           & $b$  $\bar b$  $\nu_\tau$  $\bar\nu_\tau$ & 0.021165(7) \\ 
      \hline \rule{0cm}{0.5cm}                                              
      NC32 & $s$  $\bar s$  $u$  $\bar u$ & 0.11221(8)           
           & $b$  $\bar b$  $u$  $\bar u$ & 0.09950(4) \\              
           & $d$  $\bar d$  $c$  $\bar c$ & 0.07664(4)
           & $b$  $\bar b$  $c$  $\bar c$ & 0.06284(2) \\              
      \hline \rule{0cm}{0.5cm}                                              
      NC24 & $\mu^-$  $\mu^+$  $\tau^-$  $\tau^+$ & 0.02219(3) &&\\
      \hline \rule{0cm}{0.5cm}                                              
      NC48 & $\mu^-$  $\mu^+$  $\mu^-$  $\mu^+$  & 0.01842(6)
           & $\tau^-$  $\tau^+$  $\tau^-$  $\tau^+$ & 0.005381(2) \\    
      \hline \rule{0cm}{0.5cm}                                              
      NC64 & $d$  $\bar d$  $d$  $\bar d$ & 0.03913(2)
           & $b$  $\bar b$  $b$  $\bar b$ & 0.025382(5) \\              
           & $s$  $\bar s$  $s$  $\bar s$ & 0.03608(1)
        &&\\                                             
      \hline \rule{0cm}{0.5cm}                                              
      NC24 & $\mu^-$  $\mu^+$  $d$  $\bar d$ & 0.06194(6)          
           & $\tau^-$  $\tau^+$  $d$  $\bar d$ & 0.02987(1) \\         
           & $\mu^-$  $\mu^+$  $s$  $\bar s$ & 0.06065(5)             
           & $\tau^-$  $\tau^+$  $s$  $\bar s$ & 0.02880(1) \\         
           & $\mu^-$  $\mu^+$  $b$  $\bar b$ & 0.05516(2)             
           & $\tau^-$  $\tau^+$  $b$  $\bar b$ & 0.024889(5) \\         
      \hline \rule{0cm}{0.5cm}                                              
      NC32 & $d$  $\bar d$  $s$  $\bar s$ & 0.07531(3)            
           & $s$  $\bar s$  $b$  $\bar b$ & 0.06128(2) \\              
           & $d$  $\bar d$  $b$  $\bar b$ & 0.06413(2)                
        &&\\                                             
      \hline
    \end{tabular} 
    \caption{Cross-sections for all the neutral current 4-$f$
      processes not compatible with $\gamma\gamma$-like final states
      at $\sqrt{s} = 199.5$~GeV with the phase space cuts discussed in
      the text. The errors quoted are purely statistical, corresponding to the
      accuracy of the integration.} 
    \label{tab:ncxsec}
  \end{center}
\end{table}

\begin{table}[hbtp]
  \begin{center}
    \begin{tabular}{|c|l c c|}
      \hline
      \multicolumn{4}{|c|}{} \\
      \multicolumn{4}{|c|}{\bf NC ($\gamma\gamma$ compatible)}\\
      \multicolumn{4}{|c|}{} \\
      \hline \hline
      process type & final state & 4-$f$ region cross-section (pb) &
      $\gamma\gamma$ region cross-section (pb) \\
      \hline \hline \rule{0cm}{0.5cm}                                          
      NC144 & $e^-$  $e^+$  $e^-$  $e^+$ & 2.687(9) & 5.225(6) \\
      \hline \rule{0cm}{0.5cm}                                              
      NC48 & $e^-$  $e^+$  $\mu^-$  $\mu^+$ & 1.093(2) & 21.19(2) \\        
           & $e^-$  $e^+$  $\tau^-$  $\tau^+$ & 0.5340(6) & 7.385(2)
      \\
      \hline \rule{0cm}{0.5cm}                                              
      NC48 & $e^-$  $e^+$  $u$  $\bar u$ & 1.171(2) & 10.382(4) \\           
           & $e^-$  $e^+$  $c$  $\bar c$ & 0.5761(6) & 7.378(2) \\
      \hline \rule{0cm}{0.5cm}                                              
      NC48 & $e^-$  $e^+$  $d$  $\bar d$ & 0.3682(6) & 1.6599(8) \\           
           & $e^-$  $e^+$  $s$  $\bar s$ & 0.3327(6) & 1.4825(5) \\
           & $e^-$  $e^+$  $b$  $\bar b$ & 0.24304(8) & 0.36173(9) \\
      \hline
    \end{tabular} 
    \caption{Cross-sections for the neutral current 4-$f$
      processes with $\gamma\gamma$-compatible final states, i.e. $eell$ and
      $eeqq$ where $l$ is a charged lepton and $q$ a quark, at
      $\sqrt{s} = 199.5$~GeV. The cross-section is shown separately for the two
      different phase space regions described in appendix~\ref{app1}.
      The errors quoted are purely statistical, corresponding to the
      accuracy of the integration.}
    \label{tab:ggxsec}
  \end{center}
\end{table}



\clearpage
\newpage

\begin{thebibliography}{9}
\bibitem{lep2} Physics at LEP2, G.~Altarelli, T.~Sj\"ostrand and 
 F.~Zwirner eds., CERN 96-01 (1996).
\bibitem{4fclass} D.~Bardin {\it et al.}, {\it Event Generators for
 WW Physics}, in Ref.~\cite{lep2}, vol.2 p.3 [hep-ph/9709270].
\bibitem{4fdiag} F.~Boudjema {\it et al.}, {\it Standard Model
 Processes at LEP2}, in Ref.~\cite{lep2}, vol.1 p.207 [hep-ph/9601224].
\bibitem{lep2mcws} M.~Gr\"{u}newald {\it et al.}, {\it Four-Fermion
    Production in Electron-Positron Collisions}, in {\it Report of the
  Working Groups on precision calculations for LEP2 physics}, 
  S.~Jadach, G.~Passarino and R.~Pittau eds., CERN 2000-009 (2000) 1
 [hep-ph/0005309].
\bibitem{dpa1} W.~Beenakker, F.A.~Berends and A.P.~Chapovsky,
    Nucl. Phys. {\bf B548} (1999) 3.
\bibitem{dpa2} A.~Denner, Fortschr. Phys. {\bf 41} (1993) 307; \\
A.~Denner, S.~Dittmaier and G.~Weiglein, Nucl. Phys. {\bf B440} (1995) 95; \\ 
A.~Denner, S.~Dittmaier and M.~Roth, Nucl. Phys. {\bf B519} (1998) 39.
\bibitem{RacoonWW} A.~Denner, S.~Dittmaier, M.~Roth and D.~Wackeroth,
 Nucl. Phys. {\bf B560} (1999) 33; \\
A.~Denner, S.~Dittmaier, M.~Roth and D.~Wackeroth, Nucl. Phys. {\bf
 B587} (2000) 67.  
\bibitem{Yfsww}
  S.~Jadach, W.~Placzek, M.~Skrzypek, B.~F.~L.~Ward and Z.~Was,
 Phys. Lett. {\bf B417} (1998) 326; \\
  S.~Jadach, W.~Placzek, M.~Skrzypek, B.~F.~L.~Ward and Z.~Was, Comp.
  Phys. Commun. {\bf 140} (2001) 432.  
\bibitem{dpa3} J.~Fleischer, F.~Jegerlehner and M.~Zralek,
 Z. Phys. {\bf C42} (1989) 409.
\bibitem{KCansatz} A.P.~Chapovsky and V.A.~Khoze, Eur. Phys. J. {\bf
    C9} (1999) 449.  
\bibitem{fabio} R.~Chierici and F.~Cossutti, Eur. Phys. J. {\bf C23}
  (2002) 65.  
\bibitem{KandY} S.~Jadach, W.~Placzek, M.~Skrzypek,
  B.~F.~L.~Ward and Z.~Was, Comp. Phys. Commun. {\bf 140} (2001) 475.
\bibitem{Koralw} S.~Jadach, W.~Placzek, M.~Skrzypek, B.~F.~L.~Ward and
  Z.~Was, Comp. Phys. Commun. {\bf 119} (1999) 272.  
\bibitem{wphact}
  E.~Accomando and A.~Ballestrero, Comp. Phys. Commun.
  {\bf 99} (1997) 270; \\
  E.~Accomando, A.~Ballestrero and E.~Maina, hep-ph/0204052 (2002),
 to be published in Comp. Phys. Commun.  
\bibitem{Boonekamp} M.~Boonekamp,
  DAPNIA-SPP-01-16 (2001) [hep-ph/0111213].  
\bibitem{qedps} Y.~Kurihara,
  J.~Fujimoto, T.~Munehisha and Y.~Shimizu, Progress of Theoretical
  Physics {\bf 96} (1996) 1223.  
\bibitem{ifl} E.~Accomando,
  A.~Ballestrero and E.~Maina, Phys. Lett. {\bf B479} (2000) 209.
\bibitem{pythia} T.~Sj\"{o}strand {\it et al.}, Comp.
  Phys. Commun. {\bf 135} (2001) 238.  
\bibitem{photos} E.~Barberio and
  Z.~Was, Comp.  Phys. Commun. {\bf 79} (1994) 291.  
\bibitem{tauola}
  S.~Jadach, Z.~Was, R.~Decker and J.H.~Kuehn, Comp. Phys. Commun. {\bf
    76} (1993) 361.  
\bibitem{ariadne} L.~L\"{o}nnblad, Comp. Phys.
  Commun. {\bf 71} (1992) 15.  
\bibitem{herwig} G.~Corcella {\it et al.},  JHEP {\bf 01} (2001) 10
  [hep-ph/0011363].  
\bibitem{qcd96} I.G.~Knowles {\it et al.}, {\it QCD Event
 Generators}, in Ref.~\cite{lep2}, vol.2 p.103 [hep-ph/9601212].  
\bibitem{pytuning} DELPHI Collaboration, P.~Abreu {\it et al.},
  Zeit. Phys. {\bf C73} (1996) 11.  
\bibitem{kk2f} S.~Jadach, B.~F.~L.~Ward and
  Z.~Was, Comp. Phys. Commun. {\bf 130} (2000) 260.  
\bibitem{hzha} M.L.~Mangano {\it et al.}, {\it Event Generators for Discovery
 Physics}, in Ref.~\cite{lep2}, vol.2 p.309 [hep-ph/9602203].  
\bibitem{bdkrc} F.A.~Berends, P.H.~Daverveldt and
  R.~Kleiss, Comp. Phys.  Commun. {\bf 40} (1986) 271.  
\bibitem{bdk}
  F.A.~Berends, P.H.~Daverveldt and R.~Kleiss, Comp. Phys. Commun. {\bf
    40} (1986) 285.  
\bibitem{tgcpar} G.~Gounaris {\it et al.}, {\it Triple Gauge Boson Couplings},
  in Ref.~\cite{lep2}, vol.1 p.525 [hep-ph/9601233].  
\bibitem{PDG} The Particle Data Group,
  D.~E.~Groom {\it et al.}, Eur. Phys. J. {\bf C15} (2000) 1.
\end{thebibliography}
\end{document}